\documentclass{aa}
\usepackage{graphicx}
\usepackage{natbib}
\usepackage{rotating}
\usepackage{txfonts}
\bibpunct{(}{)}{;}{a}{}{,}

\usepackage{fix2col} 

\newcommand\changed{}

\begin{document}
\title{The Polarization of Drifting Subpulses} \author{R.~T. Edwards}
\institute{Astronomical Institute ``Anton Pannekoek'', University of
Amsterdam, Kruislaan 403, 1098 SJ Amsterdam, The Netherlands}

\offprints{R.~T. Edwards,\\ \email{redwards@science.uva.nl}}
\date{Received 5 April 2004 / Accepted 12 July 2004}

\abstract{ Using new techniques based on the polarimetric fluctuation
spectrum, the fluctuation behaviour of the polarization of individual
pulses is examined in three pulsars that show drifting subpulses,
allowing various aspects of the fluctuations to be quantified for the
first time. Of the three pulsars studied, only PSR B0809+74 shows
behaviour completely consistent with the superposition of orthogonal
polarization modes (OPMs), and this only at 328 MHz and in
superposition with an apparently randomly polarized component.  The
observed periodic pattern is decomposed into the sum of two
orthogonally polarized, out-of-phase drift patterns, one of which
shows a dramatic jump in subpulse phase near the leading edge of the
pulse window, which probably relates to the phase jump earlier
reported in total intensity at 1380~MHz.  For PSR B0320+39 and PSR
B0818$-$13, considerable periodic fluctuations away from OPM
orientations are seen, a condition that also occurs in the trailing
half of the pulse in PSR B0809+74 at 1380~MHz. In some cases the
deviation is so strong that the periodic locus of the polarization
vector in the Poincar\'{e} sphere is almost circular, in contrast to
the strictly colinear states of superposed OPMs.  Several
possibilities are discussed for the physical origin of these
patterns. The similarity between the subpulse patterns in one of the
OPMs of PSR B0809+74 at 328~MHz to that of the total intensity signal
at 1380~MHz supports a picture of superposed, out of phase drift
patterns. To explain the full range of behaviour seen in the three
pulsars, it must be possible to produce at least three arbitrarily
polarized superposed patterns. While the data do not suggest a
particular approach for the empirical decomposition of patterns into
non-orthogonally polarized components, the specific, quantitative
nature of the results should provide strong constraints for
theoretically driven modelling.   } \maketitle

\section{Introduction}
In the analysis of the bewildering variety of shapes and polarizations
of individual pulses, one tends to seek order, in the hope that an
understanding of simple, repeatable phenomena might contribute to a
more complete picture of the factors involved in pulsar emission. In
this work two such phenomena are considered: the tendency of
polarization states to cluster around two orthogonal orientations
(orthogonal polarization modes; OPMs) and the periodic occurrence of
subpulses that ``drift'' across the pulse window (drifting subpulses).
Both phenomena offer potential insight into the underlying physics,
because they can be tied by simple models to the geometry of viewing
emission that is beamed along dipolar magnetic field lines, if the
polarization is oriented parallel or perpendicular to the magnetic
field \citep{rc69a,kom70,mth75,brc76}, and the drifting subpulses
correspond to emitting entities that rotate as an ensemble about the
magnetic axis \citep{rud72,wri81,ash88,es02}.

That the polarization state is related to the subpulse modulation
pattern has been known for some time. Due to insufficient instrumental
time resolution, it was initially thought that a smooth transition of
position angle occurred over the course of each subpulse
\citep{thhm71}.  It was later conclusively shown using data from four
pulsars with drifting subpulses, that the changes are generally sharp
jumps from one orthogonal state to another, with the point of
transition being related to the subpulse position rather than the
average profile \citep{mth75}.  Two of these pulsars (PSR B0031$-$07
and PSR B0809+74) show quasi-periodic drifting, with the consequence
that, for any given pulse longitude, the OPM state alternates
periodically at the period of the subpulse drifting. Using a
high-resolution digital pulsar backend and modern visualisation
technology, this picture was confirmed by \citet{rrs+02} in
observations of PSR B0809+74 at 328~MHz.  A natural interpretation of
this phenomenon in terms of the standard rotating spark ``carousel''
model of drifting subpulses \citep{rud72} is that the observed beam is
the incoherent superposition of two orthogonally polarized,
azimuthally offset carousel beams \citep{esv03,rr03}, however to date
a quantitative test of the model and determination of the required
azimuthal offset have not been performed.

Much empirical model-building has been done on the basis of the
interpretation of data in terms of the canonical geometric models of
polarization (e.g. \citealt{ran83,lm88, ran93,md99,hm01}) and drifting
subpulses (e.g. \citealt{dr99,dr01,ad01,vsrr03, gs03,rsd03}). In
practice, however, significant deviations from the models are
frequently observed. In many pulsars, the distribution of polarization
position angles in individual pulses is broadened beyond what is
expected from the instrumental noise (e.g. \citealt{scr+84}), and a
distribution with bimodal peaks separated by an angle other than
90\degr\ is often observed (e.g. \citealt{br80}). Related to this,
mean polarization profiles are generally poorly fit by the simple
geometric model \citep{ew01}. Empirical models involving
non-orthogonal modes \citep{mck03a} and superposed randomly polarized
radiation (RPR; \citealt{mck04}) have been pursued as a first-order
explanation for non-OPM behaviour, however in at least one known case
(PSR B0329+54; \citealt{es04}), the distribution of polarization
states on the Poincar\'{e} sphere shows a remarkably complicated form
indicating a more complex origin, most likely in propagation
effects. Likewise, in the case of drifting subpulses, deviations from
the predictions of simple geometry appear to be the rule rather than
the exception \citep{es03b}. It is important that these deviations are
characterised and understood if physical information is to be garnered
from observations of polarization and drifting subpulses.

In considering the deviations from model predictions in the shape of
drift bands in PSR B0320+39, \citet{esv03} arrived at the model of
offset carousel beams mentioned earlier. They also noted that, while
the beams should be {\changed roughly} orthogonally polarized for PSR
B0809+74, to explain the total intensity observations of PSR B0320+39
the beams must in this case be incompletely and/or non-orthogonally
polarized. This is due to the fact that the two beams, offset by half
the spacing of subbeams, must be of equal mean intensity near the
centre of the pulse window in order to explain the complete lack of
periodic subpulse modulation seen there through destructive
interference. The superposition of equal amounts of completely,
orthogonally polarized radiation is incompatible with the strong
polarization present in this region of the mean pulse
profile. Therefore, the true origin of polarized drifting subpulse
patterns must in general be more complicated than, or at least
different to, the model  {\changed of offset OPMs} considered by \citet{rr03}.

In this work we present an analysis of the polarization fluctuations
associated with periodic subpulse modulation in PSR B0320+39, PSR
B0818$-$13 and PSR B0809+74, with a view to quantifying not only the
OPM behaviour and the associated phase offset between the modes, but
also deviations from OPM.  The observations and analysis techniques
are described in Sect.\ \ref{sec:obsanalysis}. Sect.\
\ref{sec:results} presents the observational results, beginning with
PSR B0809+74 as its behaviour proves easiest to understand. This is
followed by the discussion (Sect.\ \ref{sec:discussion}) and
conclusions (Sect.\ \ref{sec:conclusions}).

\section{Observations and Analysis}
\label{sec:obsanalysis}
\subsection{Observations}
\label{sec:obs}
The observations used in this work were taken from the data archive
accumulated since 1999 with the PuMa pulsar backend \citep{vkv02} of
the Westerbork Synthesis Radio Telescope. Of those pulsars
known to show nearly coherent periodic drifting subpulses,
observations were available of PSR B0320+39 \citep{iks82}, PSR
B0818$-$13 \citep{rit76, la83} and PSR B0809+74 \citep{sspw70} in the
92~cm wavelength band, and for PSR B0809+74 also in the 20~cm band.
At the start of each observing session, a linearly polarized point
source was observed with the interferometric correlator, and the
relative phases of both polarization channels of each of 14 telescopes
determined. Given these phases, and the location of the source to be
observed, the signals were added in each polarization of 1--8
10~MHz-wide observing bands, and the resultant tied array signals
processed by PuMa in its digital filterbank mode, producing the four
Stokes parameters as a function of time and frequency. A full
polarization calibration is not possible due to non-linear system
components, however the quality of polarization measurements is
sufficient if the grossest source of potential error, the relative
phase of system gain in the X and Y polarizations, is determined and
corrected \citep{es04}. In principle the phase difference is removed
as a result of the telescope phase alignment performed at the start of
each observing session, however for unknown reasons this often fails
for observations in the 92 cm band, which can be noticed by the
presence of Faraday modulation in the nominal Stokes $V$, and by
comparison with previously published polarimetric pulse profiles.  For
this reason, observations in this band were calibrated for
differential gain and phase in offline processing using the technique
of \citet{es04}. Samples were then combined across all frequency
channels, including compensation for the effects of interstellar
dispersion and Faraday rotation, and the resultant time series were
divided into segments equal in length to the apparent pulse period, to
form two dimensional arrays in pulse longitude and pulse number. It is
important to note, particularly in the case of PSR B0809+74 where
observations at two frequencies were used, that the alignment of ``pulse
longitude'' to the rotational phase of the pulsar includes an arbitrary
offset.

\subsection{Analysis}
\label{sec:analysis}
For the purposes of qualitative visualisation and comparison with
earlier works, use is made use of plots of individual pulse sequences,
mean drift bands (formed by folding the data modulo the subpulse
modulation period), and histograms of the polarization
orientation. However, these techniques do not extract all of the
available information about polarization fluctuations, and to this end
we define and use techniques that are a generalisation of the
longitude-resolved fluctuation spectrum (LRFS; \citealt{bac70b}) to
the four Stokes parameters. These techniques allow for a more
sensitive and quantitative characterisation of the polarization
fluctuations.  Along with a general description of the form of
fluctuations expected under {\changed orthogonal polarization modes
(OPMs)}, polarimetric functions analogous to the LRFS were presented by
\citet{es04} for the limited purpose of removing the effects of
scintillation on aperiodic fluctuation statistics. The descriptions
are repeated here with a particular view to application of these
techniques to quasi-periodic polarization fluctuations.

The (complex) LRFS is obtained by taking
the one-dimensional discrete Fourier transforms along vectors of constant
pulse longitude:
\begin{equation}
L_{kl} = \frac{1}{N}\sum_{j=1}^{N} e^{-2\pi ijk/N} 
    \left(I_{jl} - \left<I_l\right>\right),
\end{equation}
\noindent where $i=\sqrt{-1}$, $N$ is the number of pulses, $I_{jl}$
is the total intensity (Stokes I) in pulse longitude bin $l$ of pulse
number $j$, and angle brackets denote averaging over pulse number. The
corresponding information for fluctuations in the abstract vector
space of the Poincar\'{e} sphere, $\vec{p_{jl}}=\left(Q\;U\;V\right)^T$,
is given by the ``polarization LRFS'' (PLRFS):
\begin{equation}
\vec{P_{kl}} = \frac{1}{N}\sum_{j=1}^{N} e^{-2\pi ijk/N}
\left(\vec{p_{jl}}-\left<\vec{p_l}\right>\right) .
 \label{eq:plrfs}
\end{equation}
\noindent Any sequence of $N$ integer values of $k$ provides complete
information on the fluctuation statistics, given the aliasing effected
by the finite sample interval (one pulse period). It is most
convenient to take $k$ in the range $(-N/2,N/2]$, in the knowledge
that the part of the spectrum with $k\geq0$ corresponds to the
spectrum of the analytic signal, while the fact that the input signal
is real-valued gives the result that $\vec{P_{kl}} =
\vec{P^*_{(-k)l}}$.

As with the complex LRFS, the PLRFS decomposes the signal in each
longitude bin into a sum of sinusoids. Because subpulse modulations are
not perfectly periodic, the fundamental response is not confined to a
single coefficient of the spectrum, and some means to account for this
is necessary for optimal sensitivity. The approach taken here is to
take the LRFS and PLRFS of short (128-pulse) segments of data, over
which the response is typically confined to one coefficient, and to add
together, for each longitude bin, the appropriate coefficient from each of
the spectra, after compensating them for an arbitrary,
longitude-independent phase offset. The latter is determined
iteratively using the algorithm given by \citet{esv03}. The results
are longitude-dependent complex envelopes in total intensity and
polarization.  Because the phase and amplitudes can be different in each
of the three components of $\vec{p}$, elements of the PLRFS or the
complex polarization envelope in general describe ellipses in
$\vec{p}$-space (see also, for example, \citealt{bw99}, for a general
treatment of complex 3-vectors as phased ellipses). The parameters of
the ellipse may be obtained by observing that any complex vector
$\vec{P}$ may be written in terms of the real orthogonal vectors
describing the semi-major and semi-minor axes $\vec{A}$ and $\vec{B}$
and the phase ($\phi$) at which $\vec{P}$ is measured
\begin{equation}
\vec{P} = \left(\vec{A} + i\vec{B}\right)e^{i\phi} .
\label{eq:pellipse}
\end{equation}
The phase may be determined (modulo $\pi$) from the argument of
$\vec{P}^2 \equiv \vec{P}\cdot\vec{P}$:
\begin{equation}
\vec{P}^2 = \left(\vec{A}^2 - \vec{B}^2\right)e^{i2\phi},
\end{equation}
and used to obtain $\pm\vec{A}$ and $\pm\vec{B}$  as the
real and imaginary parts of $\vec{P}e^{-i\phi}$. Since orthogonal
polarization states are anti-parallel in $\vec{p}$-space, fluctuations
due to pure OPM behaviour should be strictly colinear, i.e. $|\vec{B}|=0$.
In this case, the phases and amplitudes of modulations in the two
modes can be determined as the arguments and moduli of $m_1$ and
$m_2$,  from the relations
\begin{eqnarray}
L &=& m_1 + m_2 \\
\vec{P} &=& \vec{A}e^{i\phi} = \vec{1}_A\left(m_1 - m_2\right),
\end{eqnarray}
\noindent where $\vec{1}_A \equiv \vec{A}/|\vec{A}|$ is a unit vector
parallel to the polarization orientation of mode 1 (i.e.\ the
direction in which all fluctuations occur), $L$ and $\vec{P}$ are
elements of the LRFS and PLRFS or of the determined complex envelopes,
and the longitude dependence is dropped for clarity. These 
relations describe the incoherent superposition of radiation in two
orthogonally polarized modes,  and can be
trivially solved to give
\begin{eqnarray}
m_1 &=& \left(L + \vec{P}\cdot\vec{1}_A\right)/2 \label{eq:decomp1}\\
m_2 &=& \left(L - \vec{P}\cdot\vec{1}_A\right)/2 \label{eq:decomp2}.
\end{eqnarray}

Note that the PLRFS is the analogue of the {\it complex} LRFS.  In
situations where phase information is irrelevant, or where the pulse
sequence is to be divided into segments and the resulting spectra
added (e.g. if the desired spectral resolution is much lower than the
number of pulses available), LRF {\it power} spectra are generally
used (e.g. \citealt{bac73,es03a}). Unlike the LRFS, for the PLRFS the
analogue of the power spectrum is not simply obtained by taking the
element-wise squared magnitudes of the complex spectrum, for this
would neglect possible correlation between the three components of
$\vec{p}$. The complete autocorrelation statistics of $\vec{p}$ must
also include covariance terms, a condition satisfied by the spectral
density tensor
\begin{equation}
\vec{S_{kl}} = \vec{P_{kl}}\vec{P_{kl}}^\dag,
 \label{eq:spec}
\end{equation}
\noindent where $^\dag$ denotes the Hermitian transpose. Unlike the
PLRFS, the spectral density tensor contains no absolute phase
information, and can thus be summed over multiple segments of data
without the need to track phase from segment to segment (which in any
case is only possible for nearly coherent signals).  Viewed as a
two-dimensional array of $3\times 3$ matrices, its elements are the
coherence matrices of $\vec{p}$, for a given fluctuation frequency
($k/N$ cycles per pulse period) in a given pulse longitude bin ($l$).

Summing the tensor over a range of $k$ provides information on the
fluctuation statistics of the data over the corresponding frequency
range, with the limiting case that the sum over all $k$ gives the
real-valued covariance matrix of the data. In practice, small values
of $|k|$ from are excluded the sum to remove bias due to
scintillation-induced fluctuations, and the mean spectral density
tensor (estimated from off-pulse longitudes) is subtracted to remove
measurement noise bias \citep{es04}. The covariance matrix can be
subject to eigenvector decomposition to provide information on the
principal directions of the fluctuations and the shape of the cloud
formed in $\vec{p}$-space \citep{mck04,es04}. These are specified in
terms of an orthogonal basis composed of the three unit eigenvectors,
in combination with the three eigenvalues that give the variance of
$\vec{p}$ along the corresponding directions. The covariance of
$\vec{p}$ between components in any pair of eigenvectors is zero, thus
the decomposition effectively models the distribution of $\vec{p}$ as
an ellipsoidal cloud, with the eigenvectors as its axes.  In addition
to the individual eigenvalues, use shall be made of the polarization
entropy \citep{es04}, which combines the information from the three
eigenvalues to give a measure of how isotropic the fluctuations are,
on scale of 0 (all points colinear) to 1 (completely isotropic,
spherical distribution).  Since OPMs define antiparallel orientations
in $\vec{p}$-space, the signature of pure OPM behaviour is perfect
linear correlation, with only one significantly positive eigenvalue
(with a corresponding vector aligned with the modal orientation) and
an entropy of zero.

In principle one may also measure the covariance matrix summed over
$k\geq0$, or a subset of positive $k$, to obtain the coherence matrix
of the analytic signal, or a desired frequency interval of the
analytic signal, corresponding for example to quasi-periodic
modulations. The result could be decomposed using eigenanalysis into
the incoherent sum of orthogonal elliptical fluctuations, or analysed
using other techniques of three-dimensional polarimetry
(e.g. \citealt{sam73, ckb00b,den04}). However, the pulsars covered in
this work have sufficiently coherent fluctuations that the PLRFS and
the real (scintillation-corrected) covariance matrix provide all of
the required information on the periodic and non-periodic fluctuations
observed.

\begin{figure*}
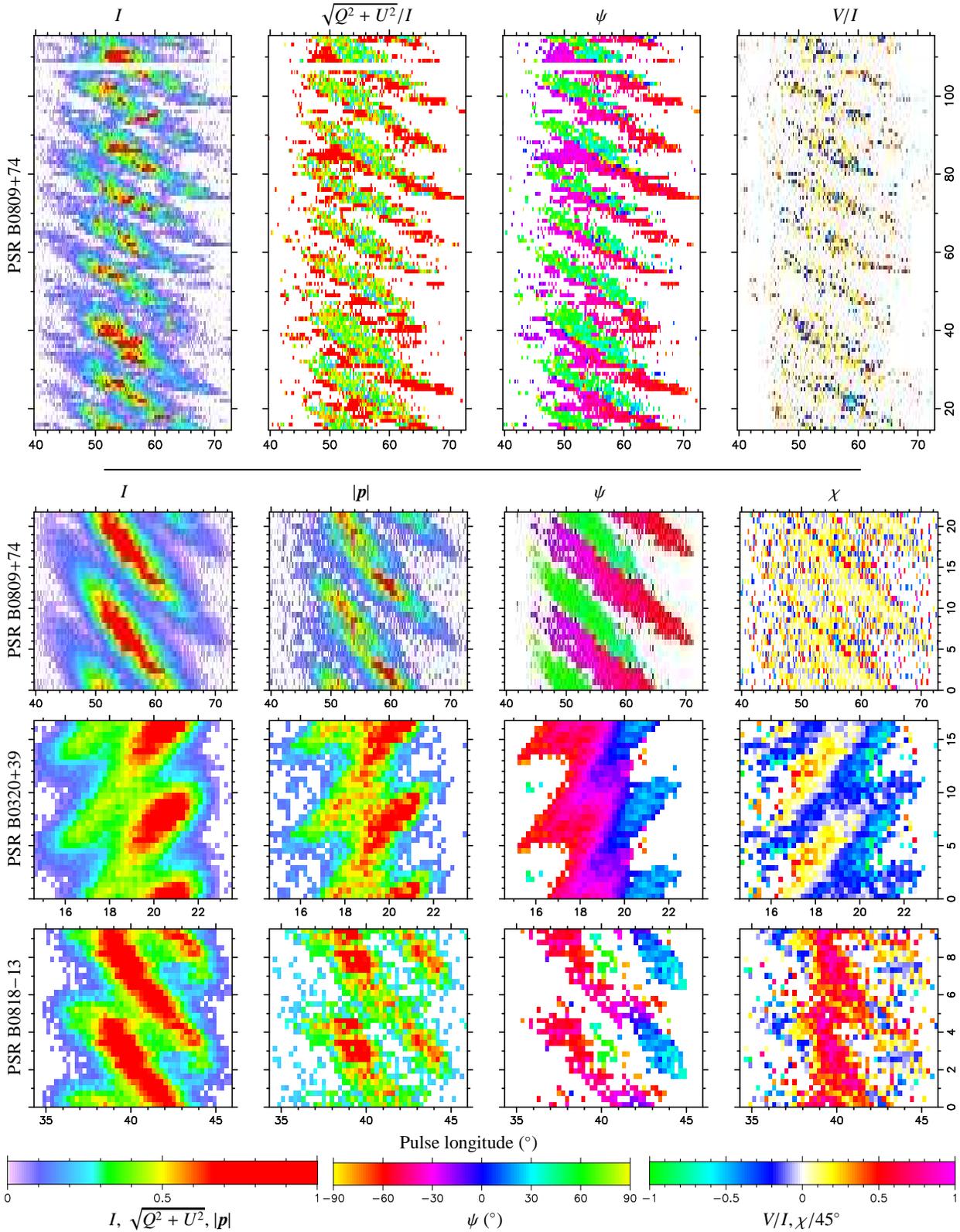

\hspace{2.6cm} $I$ \hspace{3.2cm} $\sqrt{Q^2+U^2}/I$ 
\hspace{3.0cm} $\psi$ \hspace{3.7cm} $V/I$\\
\centerline{\rotatebox{90}{\hspace{3cm} PSR B0809+74}
\resizebox{0.89\hsize}{!}{\includegraphics{0809_328seq.eps}}} 
\centerline{\parbox{13cm}{\hrulefill}}
\indent
\hspace{2.7cm} $I$ \hspace{3.7cm} $|\vec{p}|$ 
\hspace{3.7cm} $\psi$ \hspace{3.7cm} $\chi$\\
\centerline{\rotatebox{90}{\hspace{8mm} PSR B0809+74}
\resizebox{0.89\hsize}{!}{\includegraphics{0809_328folded.eps}}}
\centerline{\rotatebox{90}{\hspace{8mm} PSR B0320+39}
\resizebox{0.89\hsize}{!}{\includegraphics{0320folded.eps}}}
\centerline{\rotatebox{90}{\hspace{8mm} PSR B0818$-$13}
\resizebox{0.89\hsize}{!}{\includegraphics{0818folded.eps}}}
\indent \hspace{7.5cm} Pulse longitude (\degr) \\
\indent \hspace{0.7cm}
\resizebox{0.3\hsize}{!}{\includegraphics{gradient1.eps}}
\resizebox{0.3\hsize}{!}{\includegraphics{gradient2.eps}}
\resizebox{0.3\hsize}{!}{\includegraphics{gradient3.eps}} \\
\indent\hspace{2.5cm} $I$, $\sqrt{Q^2+U^2}$, $|\vec{p}|$
\hspace{3.9cm} $\psi$ (\degr) \hspace{4.3cm} $V/I$, $\chi/45\degr$
\caption{Colour plots of polarimetric subpulse patterns, pulse number
(ordinate) versus pulse longitude (abscissa). The top row shows a
sequence of single pulses PSR B0809+74 at 328~MHz, after
\citet{rrs+02} but now correctly calibrated. The other rows show
mean drift bands, produced by folding the Stokes parameters modulo
subpulse modulation period, and plotted twice for continuity. From top
to bottom: PSR B0809+74 at 328~MHz, PSR  B0320+39 at 357~MHz and PSR
B0818$-$13 at 328~MHz. In the folded plots, a parameterisation of the
polarization that splits it into intensity ($|\vec{p}|$) and
orientation (position angle, $|\psi|$ and ellipticity angle, 
{\changed $\chi=\frac{1}{2}\tan^{-1} V/(Q^2+U^2)^{1/2}$: see footnote to
Sect. \ref{sec:results:0809:328}})
are used to better assess consistency with OPMs. The colour scales
used are shown at the bottom. Intensity parameters (left colour scale)
are normalised by the peak value plotted in a given panel.}
\label{fig:color}
\end{figure*}

\begin{figure*}
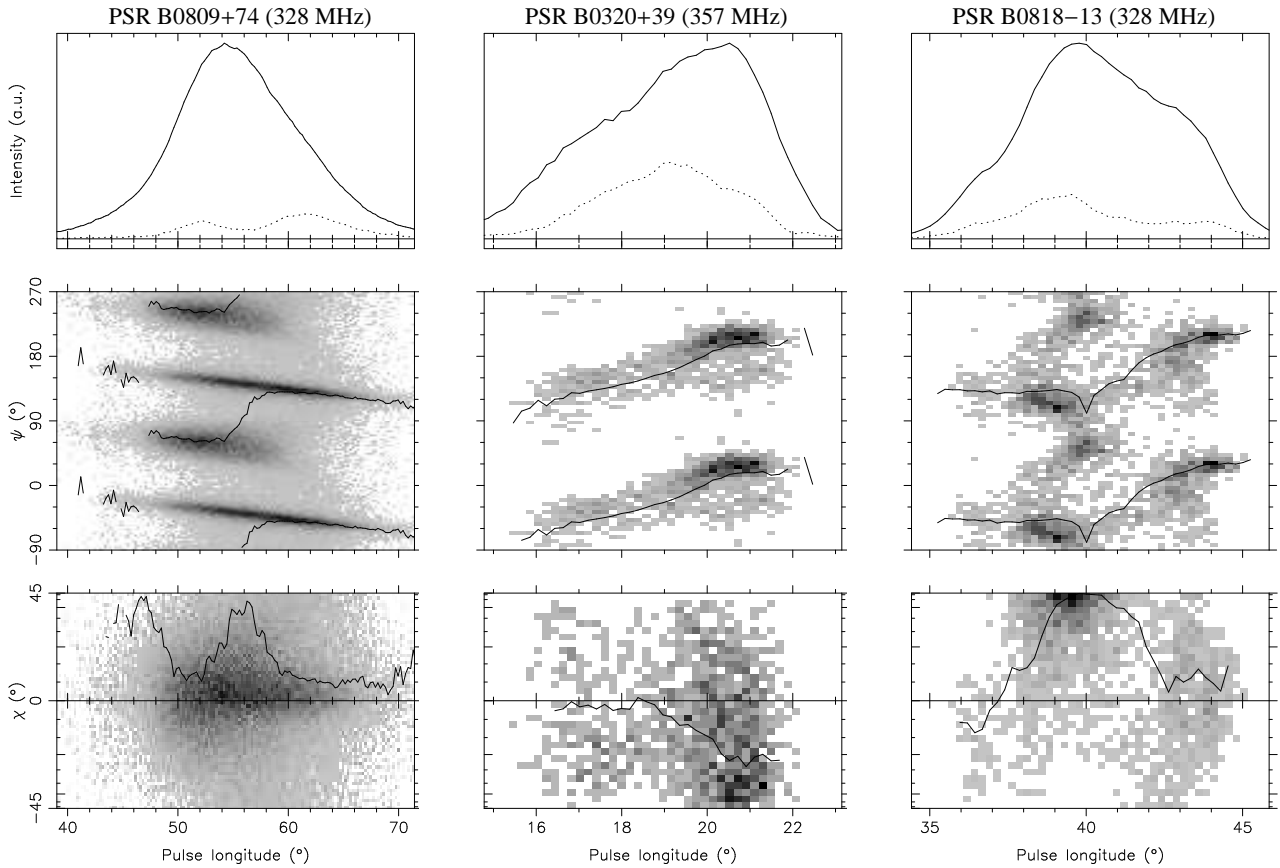

\begin{tabular}{@{\extracolsep{-5mm}}ccc}
\hspace{8mm} PSR B0809+74 (328 MHz) & 
\hspace{5mm} PSR B0320+39 (357 MHz) & 
\hspace{5mm} PSR B0818$-$13 (328 MHz) \\
\resizebox{0.32\hsize}{!}{\includegraphics{0809_328anghist.eps}} &
\resizebox{0.32\hsize}{!}{\includegraphics{0320anghist.eps}} &
\resizebox{0.32\hsize}{!}{\includegraphics{0818anghist.eps}} 
\end{tabular}
\caption{Polarization histograms.  The top panel shows the total
intensity (solid) and polarized intensity (dotted) of the average
pulse profile. The centre panel shows the longitude-dependent
histogram of position angle (plotted twice for continuity) along with
the position angle of the mean profile. The bottom panel shows the
histogram of $\sin 2\chi$ (where $\chi$ is the ellipticity angle),
along with its counterpart from the mean profile. ($\sin 2\chi$ is
chosen over $\chi$ so that each bin samples an equal area of the
Poincar\'{e} sphere; see \citealt{es04}.) For convenience the scale is
drawn (non-linearly) in terms $\chi$. Only samples with a polarized
intensity greater than three times the RMS noise (estimated from the
off-pulse) were included.  The density scale is such that bins
containing no samples are white, while other values are mapped
linearly from medium grey to black.}
\label{fig:anghist}
\end{figure*}

\begin{figure*}
\begin{tabular}{@{\extracolsep{-2mm}}cccc}
\hspace{7mm}PSR B0809+74 (328 MHz) & PSR B0809+74 (1380~MHz) &
PSR B0320+39 (357 MHz) & PSR B0818$-$13 (328 MHz) \\
\resizebox{!}{10cm}{\includegraphics{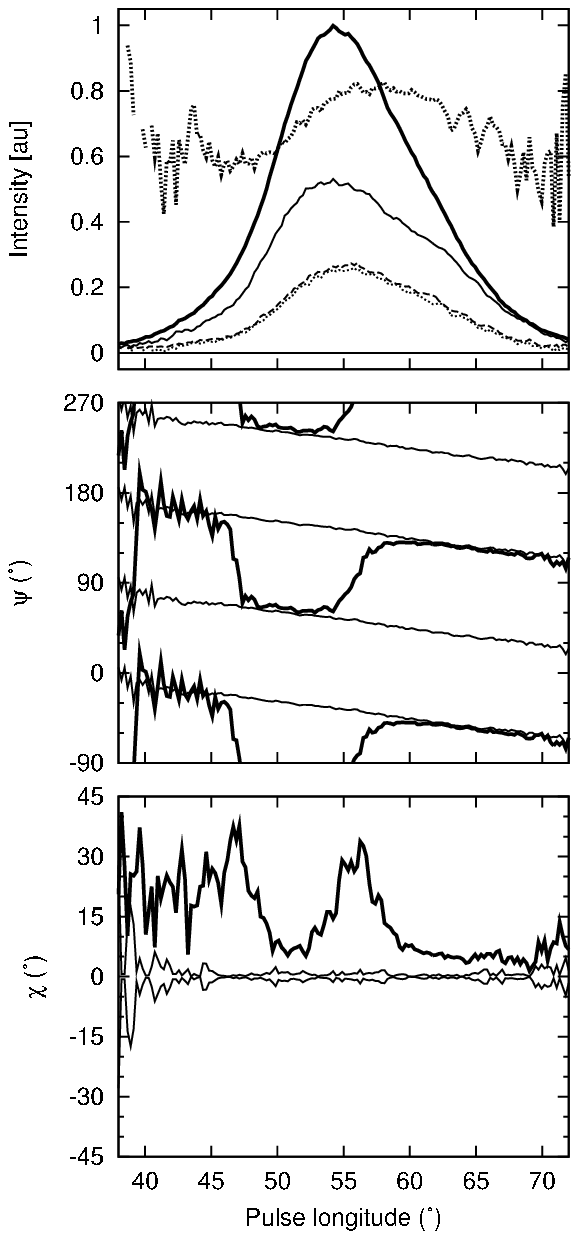}}&
\resizebox{!}{10cm}{\includegraphics{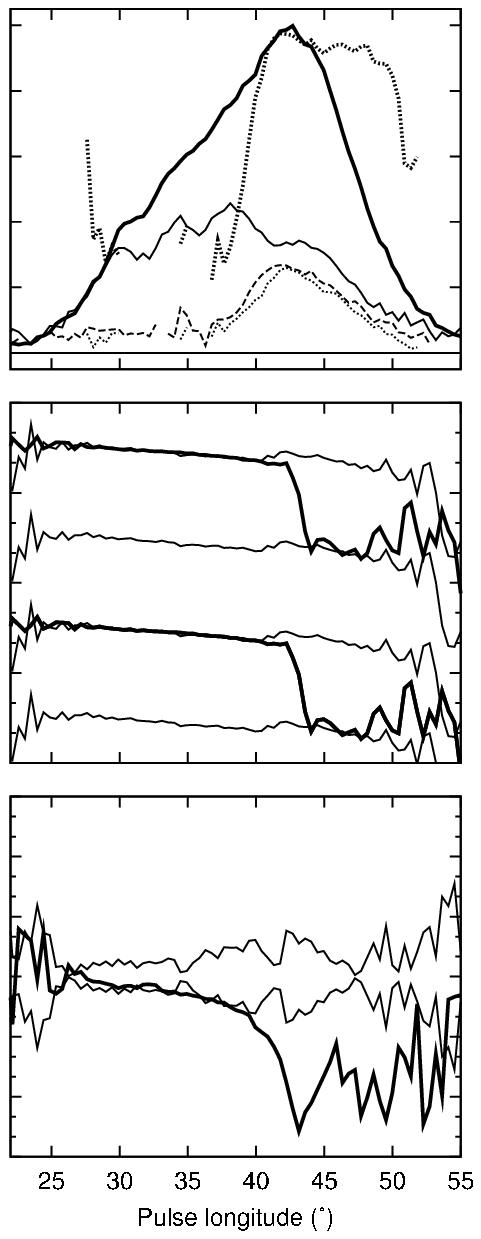}}&
\resizebox{!}{10cm}{\includegraphics{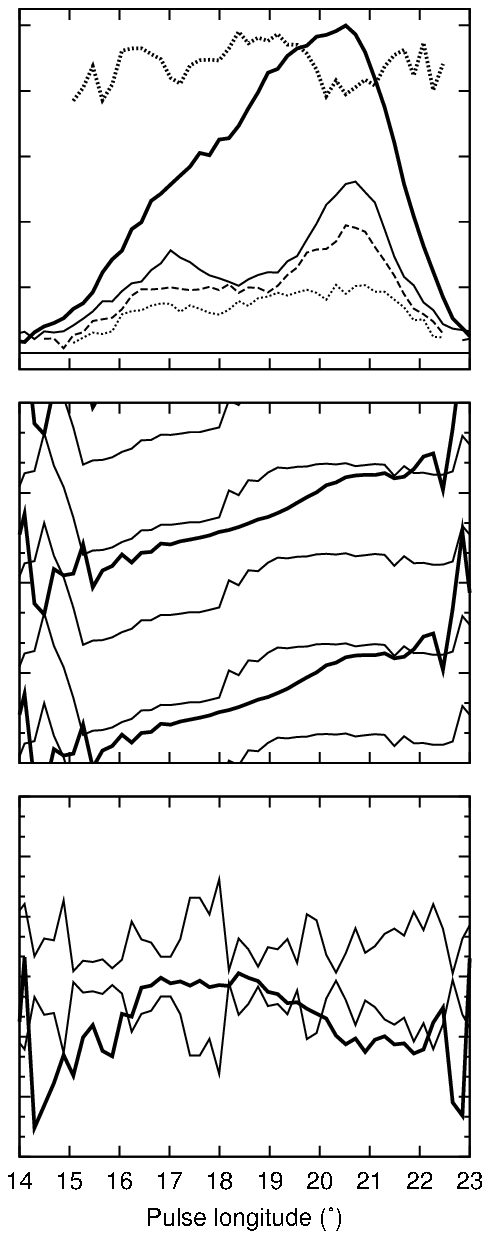}}&
\resizebox{!}{10cm}{\includegraphics{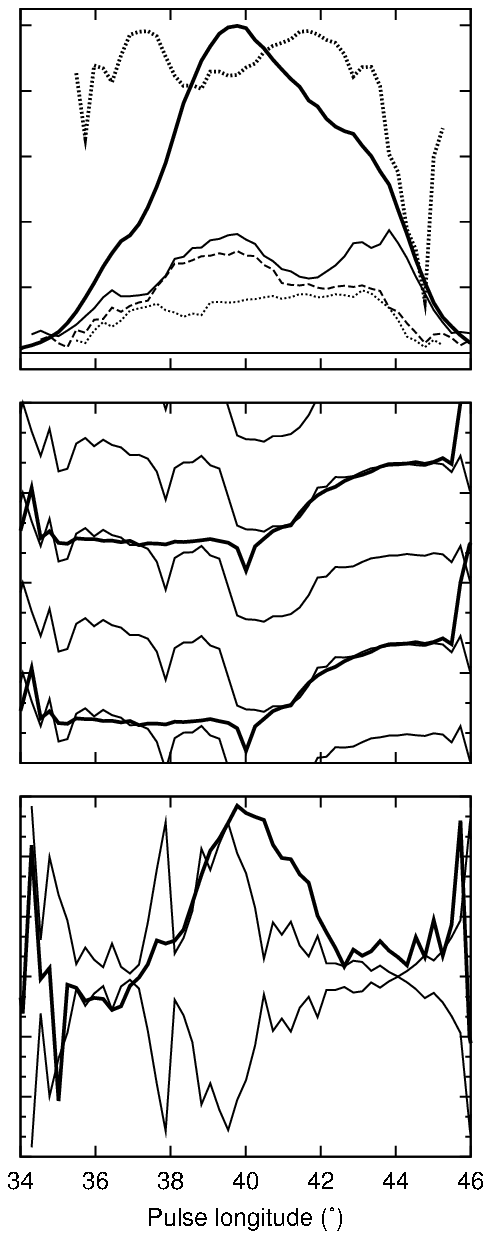}}
\end{tabular}
\caption{Results of eigenanalysis of the covariance matrices. The top
panel shows the average intensity profile (thick solid line), the
square roots of the eigenvalues (solid, dashed, dotted thin lines, in
descending order of value), and the polarization entropy (thick dotted
line). The middle panel shows the position angle of the mean
polarization vector (thick line, plotted repeatedly at offsets of
$180\degr$ for continuity) and of the first eigenvector (thin line,
plotted repeatedly at offsets of $90\degr$ for continuity under the
equivalence of antiparallel vectors as eigenvectors).  The bottom
panel shows the ellipticity angle of the mean polarization vector
(thick line) along with the ellipticity angle of the first
eigenvector, plotted twice with opposite signs.}
\label{fig:eigen}
\end{figure*}

\section{Results}
\label{sec:results}
\subsection{PSR B0809+74}
\subsubsection{328 MHz}
\label{sec:results:0809:328}
The observations of PSR B0809+74 presented here are the same ones used
by \citet{rrs+02}, who found that, at 328~MHz, each subpulse
experiences an abrupt transition between two elliptical OPMs.
Unfortunately, it appears that this observation was affected by the
the failure of the pre-observation phase calibration {\changed (Sect
\ref{sec:obs})}.  The resultant X-Y relative phase offset
(which was measured as $-51\degr$) causes a rotation of $\vec{p}$
about the Stokes $Q$ axis, resulting in mixing of $U$ and
$V$. Although this effect is complicated by the attempted correction
for Faraday rotation made before summing the frequency channels, the
net effect was that some of the linear polarization masqueraded as
circular polarization (and vice-versa). When correctly calibrated, the
pattern is qualitatively similar to that presented by \cite{rrs+02},
except that periodic sign flips appear completely absent in Stokes
$V/I$ (Fig. \ref{fig:color}), in agreement with the findings of
\citet{mth75}. The second row of Fig.\ \ref{fig:color} shows the mean
drift band, produced by folding a sequence of $\sim 300$ pulses modulo
the drift period, and plotting the result in terms of total and
polarized intensity, and polarization orientation (position angle and
ellipticity angle\footnote{The ellipticity angle
($\chi=\frac{1}{2}\tan^{-1} V/(Q^2+U^2)^{1/2}$) complements the
position angle to completely specify the orientation of a polarization
vector in $QUV$ space, independent of its length.}). This sequence was
selected as being unaffected by nulls and the associated modulation
period variations \citep{col70a,la83,vsrr03} which would smear out the
drift band shape.  Clearly seen is a reduction in the polarized
intensity in the middle of the drift band, accompanied by sharp jumps
in position angle.  No analogous jump is seen in ellipticity angle, as
would be expected for elliptical OPMs. The longitude-resolved
histograms of position and ellipticity angles (Fig.\
\ref{fig:anghist}) confirm this picture, and show that major random
deviations from the nominal linear OPM orientations occur in both
parameters.

The results of eigenanalysis of the covariance matrix (Fig.\
\ref{fig:eigen}) provide quantitative support for the features noted
above.  The very small values of ellipticity angle for the dominant
eigenmode concur with the lack of sign flips in Stokes $V/I$
(Fig. \ref{fig:color}), indicating that the OPMs are very nearly
purely linearly polarized. Incidentally, we note that such a condition
would be most unlikely under an arbitrary X-Y relative phase error
(especially given the swing in position angle with pulse longitude),
confirming the quality of the calibration. The position angle swing of
the dominant eigenvector is nearly perfectly linear, being much more
robust to slight deviations from orthogonality than the position angle
of the average profile (see also \citealt{es03b}).  The second and
third eigenvalues are both significant and nearly equal in value at
most pulse longitudes, indicating that the distribution in the
Poincar\'{e} sphere is roughly that of a prolate spheroid. This is
consistent with the superposition of an apparently randomly polarized
component on top of the OPM radiation \citep{mck04}.

To examine the relationship of OPMs with the drifting subpulse
modulation in more detail, the longitude-dependent amplitude and phase
of the quasi-periodic total intensity and polarimetric fluctuations
were computed using the methods of Sect.\ \ref{sec:analysis}. The
result is shown in Fig.\ \ref{fig:decomp}. The measured values of
$|\vec{B}|$ (semi-minor axis) are small, showing that the periodic
fluctuations in $\vec{p}$ are nearly confined to a line, consistent
with an origin in OPMs. Working from this assumption, the observed
periodicity was decomposed into the sum of two out-of-phase OPMs using
Eqs. \ref{eq:decomp1} and \ref{eq:decomp2}. The resultant
longitude-dependent amplitude and phase envelopes for the two modes
are also depicted in Fig.\ \ref{fig:decomp}, along with their phase
difference (for convenience). The amplitude peaks are offset in pulse
longitude, and for most of the pulse window the phases are offset by
$\sim 50\degr$. However, at pulse longitude $\sim 47\degr$ there is a
sharp ``jump'' of $\sim 120\degr$ in the phase envelope of one of the
modes.

\subsubsection{1380 MHz}
\label{sec:0809_1380}
Owing to the reduced depth of subpulse modulation of PSR B0809+74 at
high frequencies \citep{nuwk82}, it is difficult to obtain sensitive
measurements of its subpulse properties at 1380~MHz.  Although the
best polarimetric observation available was of significantly lower
quality than the total intensity observation used by \citet{es03b}
(which apparently encountered extraordinarily strong amplification by
the interstellar medium), some information on the subpulse behaviour
can nevertheless be obtained. The position angle histogram is similar
to the scatter plot of \citet{rrs+02}, while the ellipticity angle
histogram is broad and featureless, so for brevity these are not
displayed.

Turning to the results of eigenanalysis (Fig.\ \ref{fig:eigen}), we
see that the behaviour is very different on either side of pulse
longitude $\sim 38\degr$. On the leading side, only one significant
eigenvalue is measured, indicating very pure OPM-associated
fluctuations. On the trailing side the fluctuations become almost
isotropic, accompanied also by a $\sim 90\degr$ step in the
polarization position angle of the mean profile.

Further insight is given by the decomposition of the periodic
polarimetric fluctuations into their semi-major and semi-minor axes
(Fig. \ref{fig:decomp}). Firstly, it is seen that on the leading part
of the profile, the periodic fluctuations are approximately linear
(and of similar amplitude to those in Stokes I), while in the trailing
part significant values of $|\vec{B}|$ are detected, indicating that
the pattern cannot be produced by the superposition of two OPMs. The
phase envelope of the semi-major axis ($\vec{A}$) closely follows that
of the total intensity in the leading part of the profile, consistent
(along with the equal amplitudes) with the presence of modulations in
only one of the polarization modes. The step of $\sim 120\degr$ in the
phase envelope of the total intensity (previously reported by
\citealt{es03b}) is accompanied by a smoother, smaller transition of
the opposite sense in the phase of the semi-major axis, although the
meaning of this is not clear in the absence of a model for the origin
of elliptical periodicities.

\subsection{PSR B0320+39}
\begin{figure*}
\begin{tabular}{@{\extracolsep{-5mm}}cccc}
\hspace{6mm} PSR B0809+74 (328 MHz) & 
\hspace{3mm} PSR B0809+74 (1380~MHz) &
\hspace{3mm} PSR B0320+39 (357 MHz) & 
\hspace{3mm} PSR B0818$-$13 (328 MHz) \\
\resizebox{!}{8cm}{\includegraphics{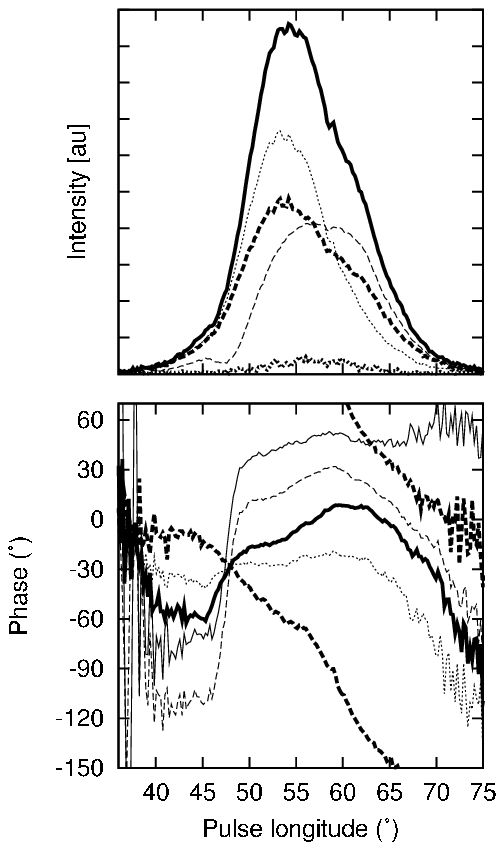}} &
\resizebox{!}{8cm}{\includegraphics{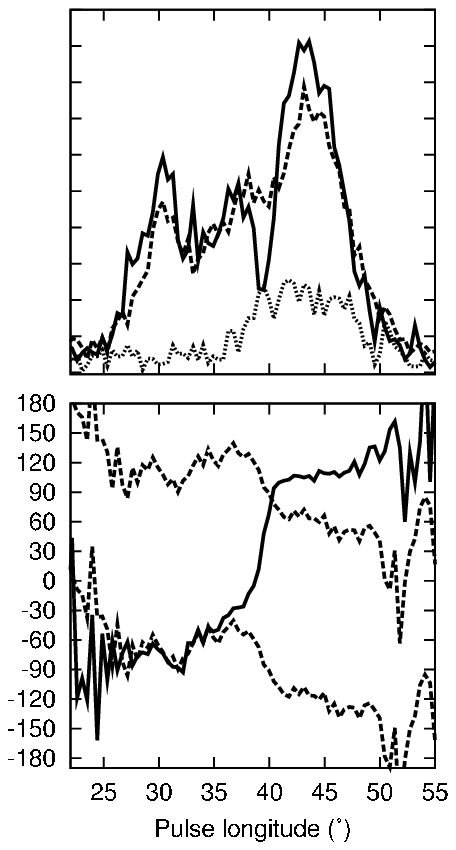}}&
\resizebox{!}{8cm}{\includegraphics{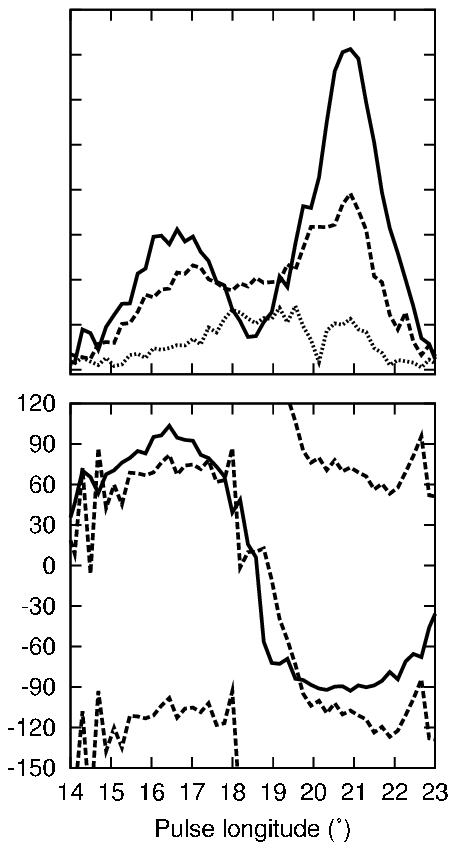}}&
\resizebox{!}{8cm}{\includegraphics{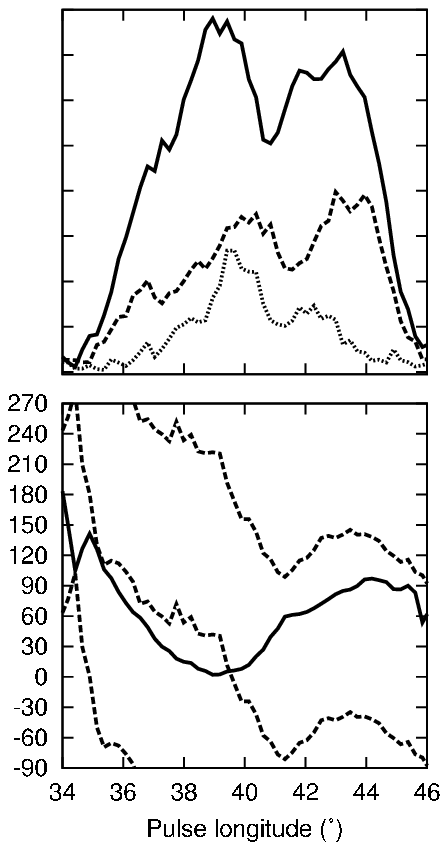}}
\end{tabular}
\caption{Amplitude (top) and phase (bottom) of subpulse modulations,
in total intensity (thick solid line) and polarization (semi-major
axis ($|\vec{A}|$): thick dashed, semi-minor axis ($|\vec{B}|$): thick
dotted). The phase of the semi-minor axis is not plotted, since by definition
it is offset from that of the semi-major axis by $\pm 90\degr$
In the left-most column only, the amplitude and phase of two OPMs
(thin dotted and dashed lines) as inferred from the total intensity
and semi-major axis values are plotted along with the phase difference
(thin solid line). A constant slope was subtracted from all phases for
clarity. The (dimensionless) slopes used were 25 for PSR B0809+74,
$-60$ for PSR B0320+39, and 60 for PSR B0818$-$13.}
\label{fig:decomp}
\end{figure*}

The orientation histograms (Fig. \ref{fig:anghist}) for PSR B0320+39
at 357~MHz show broad distributions in position angle and especially
in ellipticity angle.  This character is confirmed by the results of
eigenanalysis (Fig. \ref{fig:eigen}), which show that all three
eigenvalues are generally significant, and all different in value.
The position angle of the primary direction of fluctuation differs
markedly from that of the average profile. The amplitude and phase
envelopes of the periodic polarization fluctuations
(Fig. \ref{fig:decomp}) reveal, via the significant values for
$|\vec{B}|$, that fluctuations in at least two directions in $QUV$
space contribute to the observed periodic patterns. In fact, this can
be seen in the mean drift band (Fig.\ \ref{fig:color}): around pulse
longitudes 17--18\degr\ and 20--21\degr\ the periodicity in
ellipticity angle is clearly smoothly varying, in contrast with the
sign-reversing rectangular waveform expected under OPM.  There is some
evidence for similar behaviour in the position angle around longitude
18--20\degr.  Fig. \ref{fig:decomp} shows that semi-major axis
fluctuates roughly in phase with the total intensity, including the
$\sim 180\degr$ step in phase (longitude $\sim 18.5\degr$), which more
sensitive observations have shown is very sharp and accompanied by an
almost complete lack of periodic modulations in total intensity
\citep{esv03}. In contrast to this, significant periodic polarization
fluctuations (about an elliptical locus) are detected throughout this
part of the pulse window.

\subsection{PSR B0818$-$13}
\begin{figure}
\centerline{\resizebox{0.98\hsize}{!}{\includegraphics{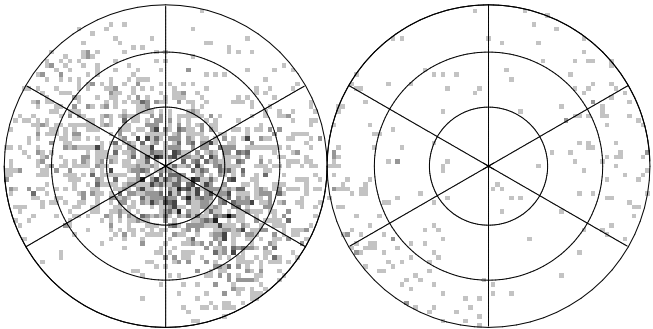}}}
\caption{Distribution of polarization orientations in pulse longitudes
39--40\degr\ for PSR B0818$-$13 at 328~MHz. The projection is
Lambert's azimuthal equal area, interrupted at the equator and
centred on the poles of the Poincar\'{e} sphere ($V/|\vec{p}| = \pm1$) :
see \citet{es04} for details. Meridians and parallels are marked
at intervals of 30\degr, with the upward vertical lines corresponding
to a position angle of 0\degr.}
\label{fig:0818pohist}
\end{figure}

The position angle histograms of PSR B0818$-$13 at 328~MHz
(Fig. \ref{fig:anghist}) reveal a chaotic jumble of preferred
polarization directions as a function of pulse longitude.  The
distributions clearly differ strongly from the expectations of OPM.  A
clue to understanding the strange behaviour around longitude
$39.5\degr$ is given in the distribution of ellipticity angles : most
points scatter about the $V/|\vec{p}|=1$ pole of the Poincar\'{e}
sphere. This is confirmed by Fig.\ \ref{fig:0818pohist}, which shows
the two-dimensional distribution on the Poincar\'{e} sphere in this
longitude region. Although the position angle distribution is roughly
bimodal, the true two-dimensional distribution is unimodal, with an
extended shape that prefers some position angles over others.  The
results of eigenanalysis (Fig.\ \ref{fig:eigen}) confirm the chaotic
picture given by the histograms, which taken together show that there
is no particular reason to consider OPMs to play a role in the
polarization fluctuations. The periodic part of the fluctuations
(Fig.\ \ref{fig:decomp}) are just as complicated, with the pattern
tracing an elliptical locus in $\vec{p}$-space, which is almost
circular at longitude $\sim 39.5\degr$ (as indicated by the near
equality of $|\vec{A}|$ and $|\vec{B}$). There is also evidence for
this in the mean drift band (Fig.\ \ref{fig:color}), in the form of
continuous variations in the ellipticity angle, and roughly bimodal
variations in the position angle.

\section{Discussion}
\label{sec:discussion}
The observations presented here prove what was already suspected in
the introduction: that periodic polarization modulations are more
complicated than anything that can be produced by the out of phase
superposition of two orthogonally polarized drift patterns (cf.\
\citealt{rr03}). Of the three pulsars studied, only PSR B0809+74 fits
this picture, and this only at 328~MHz. This is subject to the caveat
that a significant component of randomly polarized radiation (RPR)
appears to be present. 

It seems pertinent to ask whether
some other mechanism is responsible for the periodic polarization
fluctuations seen in pulsars with drifting subpulses. Whatever the
underlying mechanism for the production of drifting subpulses, it
is somehow capable of modulating the total intensity of the emission
produced in the magnetosphere. The local magnetospheric conditions
also determine the path taken by plasma wave ``rays'' in their
propagation through the magnetosphere (e.g. \citealt{fl03,pet00}), the
alterations imposed on their polarization through birefringent effects
in the polarization limiting region (e.g. \citealt{cr79,lp99}), and,
potentially, the degree of conversion to an orthogonal sense of
polarization \citep{pet01}. It therefore seems plausible that these
processes could also be modulated at the subpulse fluctuation period,
giving rise to periodic changes in the polarization state. Only
through careful modelling could the consistency of the observations
with these effects be checked.

However, there is a striking feature of the 328~MHz observations
{\changed that} is particularly suggestive that the modulations do
indeed arise due to the out of phase superposition of two or more
periodic patterns.  When the measured fluctuations were decomposed as
the sum of out of phase OPM modulations, the longitude-dependent phase
of one of the two modes was seen to execute a sudden jump of 120\degr,
accompanied by a reduction in the amplitude of fluctuations. A very
similar feature is seen in the total intensity modulations at 1380~MHz
\citep{es03b}, which, as shown in this study, occur in a single
polarization mode (at least in the leading part of the pulse
window). In our view, such a correspondence is highly unlikely to
arise unless the decomposition performed at 328~MHz has a true
physical basis. {\changed As already argued by \citet{es03b}, the
behaviour at 1380~MHz could be explained via superposed subpulse
patterns with longitude-dependent phase offsets of an arbitrary amount
(i.e. not just $180\degr$). If the same effect causes a phase jump in
a single OPM (as observed at 328~MHz), which itself is apparently
accompanied by an out-of-phase subpulse pattern in the complementary
polarization mode, then there are at least three offset subpulse
modulation components present in B0809+74. }

A second important implication of the phase jump in one OPM of PSR
B0809+74 at 328~MHz concerns the sampling of the polar cap effected in
observations at low versus high frequencies. {\changed The apparent
association of this feature with the phase jump in total intensity at
1380~MHz supports the frequency-dependent pulse profile alignment
derived from time-stamped observations by \citet{kis+98}, which
differs from the (arbitrary, visual) alignment used by
\citet{es03b}. Under the updated alignment, the leading and trailing
edges of the 1380~MHz mean profile occur significantly earlier than
the corresponding features of 328~MHz pulse, arguing that observations
at either frequency samples a region of magnetic azimuth not sampled
by the other, and/or that symmetry-breaking effects such as
abberation, retardation, refraction or asymmetric or patchy polar cap
patterns play a role. This should be contrasted to the picture of
low-frequency ``absorption'' of emission solely on the leading edge of
the profile, arrived at by \citet{bkk+81}. Their result derived from
an unphysical frequency-dependent longitude scaling performed to
explain a difference in drift rate that is actually due to the phase
step \citep{es03b}.}

The polarization fluctuations seen in PSR B0320+39 and PSR B0818$-$13
have a different character, in that the mean locus followed by the tip
of $\vec{p}$ under the periodic modulation is elliptical, rather than
linear as would be expected under the superposition of
OPMs\footnote{Purely sinusoidal modulations are confined to elliptical
paths by definition, but the presence of higher harmonics, neglected
in our analysis, could in principle allow for any closed locus in
$\vec{p}$ space. Nevertheless, the ellipticity of the fundamental is
proof that the locus is not confined to a line.}. Therefore, if the
picture of superposed, out of phase drift patterns is also to explain
the modulation in these pulsars, the component patterns must be
allowed to have polarization states that are not orthogonal (or
aligned).  That this should be the case is not surprising, for
whatever effect is responsible for the offset in modulation phase
could plausibly also affect their polarization, {\changed since both are
believed to be tied to the differences in effective viewing geometry.
Indeed, such considerations could potentially explain the slight
non-orthogonality observed even at 328~MHz in PSR B0809+74.}

\begin{figure}
\centerline{\resizebox{0.8\hsize}{!}{\includegraphics{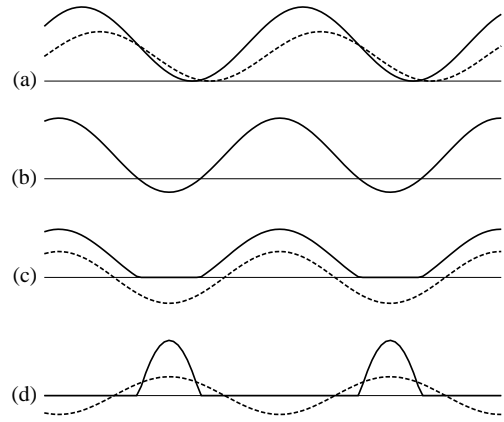}}}
\caption{Illustration of the results of three-way polarization mode
segregation \citep{dr01} on the superposition of sinusoidal OPMs.  (a)
Simulated intensity of two OPMs (in a given longitude bin) as a
function of time. (b) Resultant signal in the component of $\vec{p}$
corresponding to the modal orientations. (c) Primary polarization mode
intensity signal as inferred by three-way segregation (solid) and its
fundamental Fourier component (dashed). (d) As for (c) but for
secondary mode. Note that the underlying OPMs are nearly in phase,
while the fundamentals of the inferred signals are in pure
anti-phase.}
\label{fig:3way}
\end{figure}

Further insight into the phenomenon of superposed drift modes is
likely to come only with a larger sample size. Two previous
measurements are of relavance here. \citet{dr01} found that subbeam
patterns in two OPM-segregated polar cap images of PSR B0943+10 were
offset by roughly half the subbeam spacing.  Similarly, \citet{rr03}
found that the phases of the fundamentals in the {\changed
longitude-resolved fluctuation spectra} of the two OPMs of PSR
B1237+25 were offset by roughly 180\degr. Unfortunately, the disjoint
mode segregation algorithm used in both measurements is prone to
produce anti-phase modulation even if the patterns arise from the
nearly in-phase superposition of OPMs, since the modal patterns cannot
overlap in the segregated signal. It is easily shown (Fig.\
\ref{fig:3way}) that in the pathological case of superposed sinusoidal
modulations, the periodicity will be in complete anti-phase between
the two inferred modes, regardless of the actual phase offset. For an
accurate measurement, firstly it must ascertained whether the data are
indeed quanitatively consistent with pure OPM, and secondly the mode
separation method must take the superposition of radiation into
account.

Unrelated to the phenomenon of drifting subpulses, the 1380~MHz
results for PSR B0809+74 allow for some interesting constraints to be
placed on the origin of {\changed randomly polarized radiation (RPR)},
in light of the fact that it only occurs in the trailing half of the
pulse window {\changed (Fig. \ref{fig:eigen})}.  That this should be
accompanied by sudden, almost complete depolarization
\citep{mgs+81,rrs+02} can be understood in one of two ways. One
possibility is that the depolarization occurs due to the superposition
of two OPMs of nearly equal intensity, with the RPR being associated
with only one of the modes. The second possibility is that the RPR is
not a superposed emission component, but rather arises due to a
propagation effect, randomising the polarization of incoming radiation
and causing depolarization of the average profile. It should be noted
that this is different to the types of depolarising propagation
effects usually encountered, where the electric field vector is
altered on timescales comparable to the inverse of the observing
frequency.

The signature of RPR is significant polarization at an
orientation that is consistent over some time scale longer than the
averaging interval of the individual data samples, yet randomly
varying on some longer time scale (in our case, the pulse period). In
the case that the RPR is due to a propagation effect, this would imply
a variable transfer function (Jones or Mueller matrix) with a
decorrelation timescale that lies between the two timescales
mentioned. \citet{mck04} also mentions the viability of RPR as a
depolarization mechanism, however he appears to assume that it must
arise as superposed emission. Such a condition is not capable of
reducing the polarized intensity, as is required for PSR B0809+74.
Rather, just like unpolarized radiation, it can only reduce the {\it
fractional} polarization, by increasing the total intensity without
adding polarized intensity to the average profile. \citet{mck04}
considers the possibility of stochastic Faraday rotation in the pulsar
magnetosphere as an origin of RPR, but rejects it because the magnetic
field is unlikely to vary on short time scales, and in any case the
effect cannot alter the Stokes $V$ component of \vec{p}.  However,
this is an inadequate basis on which to rule out an origin for RPR in
propagation effects: birefringent effects in the magnetosphere can
alter all of the Stokes parameters, as the propagation modes are
believed to be linearly or elliptically polarized (rather than
circularly, as is the case for Faraday rotation), and the fluctuations
could be due to variations in plasma density and/or propagation path,
rather than magnetic field (\citealt{pet01} and references therein).

\section{Summary and Conclusions}
\label{sec:conclusions}
We have presented a detailed study of the average periodic
polarization fluctuations in three pulsars with drifting
subpulses. The basic result of periodic switching between two OPMs in
PSR B0809+74 \citep{mth75,rrs+02} is confirmed at 328~MHz, albeit in
superposition with an apparently randomly polarized component. The
periodic pattern can be understood as the superposition of drift
patterns in two orthogonal polarization modes that are offset by $\sim
50 \degr$ in subpulse phase. The drift bands are roughly linear,
however for one of the modes a sudden jump of $120\degr$, accompanied
by a reduction in the amplitude of modulations, is seen near the
leading edge of the pulse profile. Very similar behaviour is seen in
the total intensity modulations at 1380~MHz \citep{es03b}, which are
probably composed of a single polarization mode. That such a
correspondence is observed strongly supports the physical validity of
the decomposition of the pattern as the superposition of out of phase
drift patterns. Aligning the pulse profiles on the basis of the phase
jump agrees with results derived from time-stamped observations
\citep{kis+98}, implying either that different regions of magnetic
azimuth are sampled at the two frequencies, with neither frequency
presenting a complete image of the polar cap, or that other effects
such as a non-dipolar magnetic field, refraction, aberration and/or
retardation are important.

The results from PSR B0320+39 and PSR B0818$-$13 reveal that the
behaviour deviates considerably from the predictions of OPM, tracing
in general a quasi-elliptical path in the Poincar\'{e} sphere with
each cycle of the subpulse modulation. In PSR B0818$-$13, the
unimodal, asymmetric clustering of states around the $V/|\vec{p}|=1$
pole of the Poincar\`{e} sphere leads to a deceptively complicated
position angle histogram. 

In order to explain the patterns seen in the three pulsars studied,
considerable complications to the standard models are required, with
in general no less than three offset, arbitrarily polarized drift
patterns seen in superposition. For the non-orthogonal oscillations,
no particular empirical basis for decomposition is suggested by the
data, however it is hoped that the specific, quantitative nature of
the results will prove highly constraining to theoretically-driven
models.

\acknowledgements I thank B. Stappers and W. van Straten for helpful
comments on the text. The author is supported by a NOVA fellowship.
The Westerbork Synthesis Radio Telescope is administered by ASTRON
with support from the Netherlands Organisation for Scientific Research
(NWO).

\bibliographystyle{aa}

\begin{thebibliography}{52}
\expandafter\ifx\csname natexlab\endcsname\relax\def\natexlab#1{#1}\fi

\bibitem[{{Asgekar} \& {Deshpande}(2001)}]{ad01}
{Asgekar}, A. \& {Deshpande}, A.~A. 2001, MNRAS, 326, 1249

\bibitem[{{Ashworth}(1988)}]{ash88}
{Ashworth}, M. 1988, MNRAS, 230, 87

\bibitem[{Backer(1970)}]{bac70b}
Backer, D.~C. 1970, Nature, 227, 692

\bibitem[{Backer(1973)}]{bac73}
---. 1973, ApJ, 182, 245

\bibitem[{Backer \& Rankin(1980)}]{br80}
Backer, D.~C. \& Rankin, J.~M. 1980, ApJS, 42, 143

\bibitem[{{Backer} {et~al.}(1976){Backer}, {Rankin}, \& {Campbell}}]{brc76}
{Backer}, D.~C., {Rankin}, J.~M., \& {Campbell}, D.~B. 1976, Nature, 263, 202

\bibitem[{Bartel {et~al.}(1981)Bartel, Kardashev, Kuzmin, Popov, Sieber,
  Smirnova, Soglasnov, \& Wielebinski}]{bkk+81}
Bartel, N., Kardashev, N.~S., Kuzmin, A. D.~Nikolaev, N.~Y., {et~al.} 1981,
  A\&A, 93, 85

\bibitem[{Born \& Wolf(1999)}]{bw99}
Born, M. \& Wolf, E. 1999, Principles of Optics, 7th edn. (Cambridge University
  Press)

\bibitem[{Carozzi {et~al.}(2000)Carozzi, Karlsson, \& Bergman}]{ckb00b}
Carozzi, T., Karlsson, R., \& Bergman, J. 2000, Phys. Rev. E, 61, 2024

\bibitem[{Cheng \& Ruderman(1979)}]{cr79}
Cheng, A.~F. \& Ruderman, M. 1979, ApJ, 229, 348

\bibitem[{Cole(1970)}]{col70a}
Cole, T.~W. 1970, Nature, 227, 788

\bibitem[{Dennis(2004)}]{den04}
Dennis, M.~R. 2004, J. Opt. A: Pure Appl. Opt., 6, S26

\bibitem[{{Deshpande} \& {Rankin}(1999)}]{dr99}
{Deshpande}, A.~A. \& {Rankin}, J.~M. 1999, ApJ, 524, 1008

\bibitem[{Deshpande \& Rankin(2001)}]{dr01}
Deshpande, A.~A. \& Rankin, J.~M. 2001, MNRAS, 322, 438

\bibitem[{{Edwards} \& {Stappers}(2002)}]{es02}
{Edwards}, R.~T. \& {Stappers}, B.~W. 2002, A\&A, 393, 733

\bibitem[{{Edwards} \& {Stappers}(2003{\natexlab{a}})}]{es03a}
---. 2003{\natexlab{a}}, A\&A, 407, 273

\bibitem[{{Edwards} \& {Stappers}(2003{\natexlab{b}})}]{es03b}
---. 2003{\natexlab{b}}, A\&A, 410, 961

\bibitem[{{Edwards} \& {Stappers}(2004)}]{es04}
---. 2004, A\&A, 421, 681

\bibitem[{{Edwards} {et~al.}(2003){Edwards}, {Stappers}, \& {van
  Leeuwen}}]{esv03}
{Edwards}, R.~T., {Stappers}, B.~W., \& {van Leeuwen}, A.~G.~J. 2003, A\&A,
  402, 321

\bibitem[{{Everett} \& {Weisberg}(2001)}]{ew01}
{Everett}, J.~E. \& {Weisberg}, J.~M. 2001, ApJ, 553, 341

\bibitem[{Fussell \& Luo(2004)}]{fl03}
Fussell, D. \& Luo, Q. 2004, MNRAS, 349, 1019

\bibitem[{{Gil} \& {Sendyk}(2003)}]{gs03}
{Gil}, J.~A. \& {Sendyk}, M. 2003, ApJ, 585, 453

\bibitem[{{Han} \& {Manchester}(2001)}]{hm01}
{Han}, J.~L. \& {Manchester}, R.~N. 2001, MNRAS, 320, L35

\bibitem[{Izvekova {et~al.}(1982)Izvekova, Kuz'min, \& Shitov}]{iks82}
Izvekova, V.~A., Kuz'min, A.~D., \& Shitov, Y.~P. 1982, Sov. Astron., 26, 324

\bibitem[{Komesaroff(1970)}]{kom70}
Komesaroff, M.~M. 1970, Nature, 225, 612

\bibitem[{Kuzmin {et~al.}(1998)Kuzmin, Izvekova, Shitov, Sieber, Jessner,
  Wielebinski, Lyne, \& Smith}]{kis+98}
Kuzmin, A.~D., Izvekova, V.~A., Shitov, Y.~P., {et~al.} 1998, A\&AS, 127, 255

\bibitem[{Lyne \& Ashworth(1983)}]{la83}
Lyne, A.~G. \& Ashworth, M. 1983, MNRAS, 204, 519

\bibitem[{Lyne \& Manchester(1988)}]{lm88}
Lyne, A.~G. \& Manchester, R.~N. 1988, MNRAS, 234, 477

\bibitem[{Lyubarskii \& Petrova(1999)}]{lp99}
Lyubarskii, Y.~E. \& Petrova, S.~A. 1999, Astrophys. Space Sci., 262, 379

\bibitem[{Manchester {et~al.}(1975)Manchester, Taylor, \& Huguenin}]{mth75}
Manchester, R.~N., Taylor, J.~H., \& Huguenin, G.~R. 1975, ApJ, 196, 83

\bibitem[{{McKinnon}(2003)}]{mck03a}
{McKinnon}, M.~M. 2003, ApJ, 590, 1026

\bibitem[{{McKinnon}(2004)}]{mck04}
---. 2004, ApJ, 606, 1154

\bibitem[{Mitra \& Deshpande(1999)}]{md99}
Mitra, D. \& Deshpande, A.~A. 1999, A\&A, 346, 906

\bibitem[{Morris {et~al.}(1981)Morris, Graham, Seiber, Bartel, \&
  Thomasson}]{mgs+81}
Morris, D., Graham, D.~A., Seiber, W., Bartel, N., \& Thomasson, P. 1981,
  A\&AS, 46, 421

\bibitem[{Nowakowski {et~al.}(1982)Nowakowski, Usowicz, Wolszczan, \&
  K\c{e}pa}]{nuwk82}
Nowakowski, L., Usowicz, J., Wolszczan, A., \& K\c{e}pa, A. 1982, A\&A, 116,
  158

\bibitem[{{Petrova}(2000)}]{pet00}
{Petrova}, S.~A. 2000, A\&A, 360, 592

\bibitem[{{Petrova}(2001)}]{pet01}
---. 2001, A\&A, 378, 883

\bibitem[{Radhakrishnan \& Cooke(1969)}]{rc69a}
Radhakrishnan, V. \& Cooke, D.~J. 1969, Astrophys. Lett., 3, 225

\bibitem[{{Ramachandran} {et~al.}(2002){Ramachandran}, {Rankin}, {Stappers},
  {Kouwenhoven}, \& {van Leeuwen}}]{rrs+02}
{Ramachandran}, R., {Rankin}, J.~M., {Stappers}, B.~W., {Kouwenhoven},
  M.~L.~A., \& {van Leeuwen}, A.~G.~J. 2002, A\&A, 381, 993

\bibitem[{Rankin(1983)}]{ran83}
Rankin, J.~M. 1983, ApJ, 274, 333

\bibitem[{Rankin(1993)}]{ran93}
---. 1993, ApJ, 405, 285

\bibitem[{{Rankin} \& {Ramachandran}(2003)}]{rr03}
{Rankin}, J.~M. \& {Ramachandran}, R. 2003, ApJ, 590, 411

\bibitem[{{Rankin} {et~al.}(2003){Rankin}, {Suleymanova}, \&
  {Deshpande}}]{rsd03}
{Rankin}, J.~M., {Suleymanova}, S.~A., \& {Deshpande}, A.~A. 2003, MNRAS, 340,
  1076

\bibitem[{Ritchings(1976)}]{rit76}
Ritchings, R.~T. 1976, MNRAS, 176, 249

\bibitem[{Ruderman(1972)}]{rud72}
Ruderman, M. 1972, Ann. Rev. Astr. Ap., 10, 427

\bibitem[{Samson(1973)}]{sam73}
Samson, J.~C. 1973, Geophys. J. R. astr. soc., 34, 403

\bibitem[{Stinebring {et~al.}(1984)Stinebring, Cordes, Rankin, Weisberg, \&
  Boriakoff}]{scr+84}
Stinebring, D.~R., Cordes, J.~M., Rankin, J.~M., Weisberg, J.~M., \& Boriakoff,
  V. 1984, ApJS, 55, 247

\bibitem[{Sutton {et~al.}(1970)Sutton, Staelin, R.M., \& Weimer}]{sspw70}
Sutton, J.~M., Staelin, D.~H., R.M., P., \& Weimer, R. 1970, ApJ, 159, L89

\bibitem[{Taylor {et~al.}(1971)Taylor, Huguenin, Hirsch, \&
  Manchester}]{thhm71}
Taylor, J.~H., Huguenin, G.~R., Hirsch, R.~M., \& Manchester, R.~N. 1971,
  Astrophys. Lett., 9, 205

\bibitem[{{van Leeuwen} {et~al.}(2003){van Leeuwen}, {Stappers},
  {Ramachandran}, \& {Rankin}}]{vsrr03}
{van Leeuwen}, A.~G.~J., {Stappers}, B.~W., {Ramachandran}, R., \& {Rankin},
  J.~M. 2003, A\&A, 399, 223

\bibitem[{{Vo{\^ u}te} {et~al.}(2002){Vo{\^ u}te}, {Kouwenhoven}, {van Haren},
  {Langerak}, {Stappers}, {Driesens}, {Ramachandran}, \& {Beijaard}}]{vkv02}
{Vo{\^ u}te}, J.~L.~L., {Kouwenhoven}, M.~L.~A., {van Haren}, P.~C., {et~al.}
  2002, A\&A, 385, 733

\bibitem[{{Wright}(1981)}]{wri81}
{Wright}, G.~A.~E. 1981, MNRAS, 196, 153

\end{thebibliography}

\end{document}